\documentclass[sigconf,nonacm]{acmart}
\usepackage{multirow}
\usepackage{array}
\usepackage{booktabs}
\usepackage{amsmath}
\usepackage{xspace}

\usepackage{xcolor}
\usepackage{colortbl}
\usepackage{soul}
\usepackage{float}

\newcommand{\colorunderline}[2]{{\setulcolor{#1}\ul{#2}}}

\definecolor{lightblue}{RGB}{245, 47, 97}
\definecolor{lightgreen}{RGB}{245, 196, 47}
\definecolor{lightpink}{RGB}{194, 47, 245}

\newcommand{\catlabel}[1]{\setulcolor{lightgreen}\ul{#1}}
\newcommand{\rolelabel}[1]{\setulcolor{lightblue}\ul{#1}}
\newcommand{\litigantlabel}[1]{\setulcolor{lightpink}\ul{#1}}

\newcommand{\ad}[1]{\textbf{\colorunderline{lightgreen}{#1}}}
\newcommand{\role}[1]{\textbf{\colorunderline{lightblue}{#1}}}
\newcommand{\litigant}[1]{\textbf{\colorunderline{lightpink}{#1}}}

\newcommand{\adtext}[1]{\colorbox{lightgreen}{#1}}
\newcommand{\roletext}[1]{\colorbox{lightblue}{#1}}
\newcommand{\litiganttext}[1]{\colorbox{lightpink}{#1}}

\newcommand{\case}[1]{\textit{#1}}

\newcommand{\ip}{\catlabel{Intellectual Property}\xspace}
\newcommand{\antitrust}{\catlabel{Anti-competition}\xspace}
\newcommand{\dataprotection}{\catlabel{Data Protection}\xspace}
\newcommand{\tort}{\catlabel{Harmful Conduct}\xspace}
\newcommand{\justice}{\catlabel{Justice and Equity}\xspace}
\newcommand{\consumerprotection}{\catlabel{Consumer Protection}\xspace}
\newcommand{\ailegal}{\catlabel{AI in Legal Proceedings}\xspace}

\newcommand{\general}{\rolelabel{General AI Technology}\xspace}
\newcommand{\genai}{\rolelabel{Generative AI}\xspace}
\newcommand{\bio}{\rolelabel{Biometric identification technologies}\xspace}
\newcommand{\rec}{\rolelabel{Recommendation and ranking algorithms}\xspace}
\newcommand{\decision}{\rolelabel{Predictive and decision-making systems}\xspace}
\newcommand{\compvis}{\rolelabel{Virtual reality and vision technologies}\xspace}

\usepackage{newfloat}
\usepackage{listings}
\DeclareCaptionStyle{ruled}{labelfont=normalfont,labelsep=colon,strut=off} % DO NOT CHANGE THIS
\title{Visible to the Court: How AI Is (and Isn't) Litigated in U.S. Federal Court Opinions}
\author{Julie Yu}
\affiliation{%
  \institution{University of Washington}
  \city{Seattle}
  \state{WA}
  \country{USA}}
\email{juliey12@uw.edu}

\author{Rock Yuren Pang}
\affiliation{%
  \institution{University of Washington}
  \city{Seattle}
  \state{WA}
  \country{USA}}
\email{ypang2@cs.washington.edu}

\author{Jevan Hutson}
\affiliation{%
  \institution{University of Washington}
  \city{Seattle}
  \state{WA}
  \country{USA}}
\email{jevanh@uw.edu}

\author{Katharina Reinecke}
\affiliation{%
  \institution{University of Washington}
  \city{Seattle}
  \state{WA}
  \country{USA}}
\email{reinecke@cs.washington.edu}

\begin{document}

\begin{abstract}
In the United States, artificial intelligence (AI) is rapidly deployed amid limited federal regulation. With courts become a recurring forum in which AI-related practices are scrutinized, %, shaping how AI-related conduct and harms become legally legible and actionable.
it is important to empirically understand the AI litigation landscape to date. We address this gap through a systematic review of 559 U.S. federal court opinions in which AI plays a role in the parties’ contentions, 
taxonomizing (1) common topics of dispute, (2) the AI technologies implicated, and (3) the parties involved, including common plaintiff and defendant types. We identify seven recurring dispute areas, six categories of AI technologies at the center of litigation, and four types of common litigants, alongside legal doctrines used by the litigants. 
A comparison of this taxonomy to the AI Incident Database revealed substantial gaps in coverage, definitions, and prevalence between documented and litigated harms, suggesting courts capture only part of the AI risk landscape.
In addition, we found that court decisions primarily rely on pre-existing legal doctrines to manage AI rather than making new AI-specific laws, producing a form of ``piecemeal'' AI governance. 
As a result, federal court outcomes are shaped less by where AI has caused harms and more by which harms are cognizable under existing statutes, leading to certain AI harms remaining unresolved.
\end{abstract}
\maketitle


\section{Introduction}

There has been increasing public concern regarding the societal impact of artificial intelligence (AI) systems in the United States, with reported AI-related incidents %(e.g., AI-manipulated images causing defamation~\cite{mcgregor2021preventing})  
 increasing annually by 20\%-30\%~\cite{maslej2024artificialintelligenceindexreport}.

Despite regulatory efforts in some parts of the world~\cite{10.1145/3715275.3732059, ulnicane2022artificial} and attempts by U.S. state legislatures~\cite{10.1145/3657054.3657148, migineishvili2026regulating}, the U.S. has not established a federal regulatory framework for AI use. Recent executive orders rescinded earlier AI safeguards and instructed various federal agencies to fast-track AI adoption under a less centrally regulated framework~\cite{executive2025removing, executive2025removing2}. 
In this vacuum, questions about AI risks and accountability are increasingly addressed through the courts, shaping expectations on AI development and use~\cite{kaminski2021right, yew2022regulating}. 
The judiciary serves as an accountability system for harms \cite{10.1162/DAED_a_00295}, addressing issues such as copyright disputes (e.g., New York Times v. OpenAI case) and data protection issues (e.g., Clearview AI Consumer Privacy). This proliferation of AI-related litigation reflects the technology's expanding role in everyday life, and the growing tensions it generates~\cite{maslej2023artificialintelligenceindexreport}.

While high-profile cases may attract media attention, there is limited empirical understanding of what 
%the primary issues are that are being 
primary issues are
contested in courts, and what broader litigation patterns exist.
What issues are commonly litigated in court, and are these AI-related court cases increasing in frequency? Which AI technologies are the subject of conflict? Who is filing the cases, and against whom? And do the cases represent observed AI risks? Knowing the answer to these questions is critical for many stakeholders, including the public, lawyers, and policymakers
%, who may refer to specific precedents when crafting regulations
.
We answered these questions by systematically analyzing 559 federal court opinions where AI plays a role (filed between 2006 and 2026), which we retrieved from the U.S. Government Publishing Office (GovInfo.gov), an official repository providing public access to federal court opinions. 
%
%We chose court opinions as they are the primary source of what issues parties are contesting~\cite{FJCJudicialWritingManual, CornellWexOpinion, HallWright2008}, in line with our goal to understand the contested AI legal issues in court. 
% 
%
%We conducted a systematic analysis of all U.S. federal opinions in the GovInfo database in which AI plays a role, analyzing 559 relevant cases that were filed between 2006 and 2026 
%
We contribute: 
\begin{enumerate}
\item We identify seven categories of legal contention in U.S. court opinions involving AI, including Anti-competition, Intellectual Property, Data Protection, Harmful Conduct, Justice and Equity, Consumer Protection, and AI in Legal Proceedings.%\footnote{An opinion can have multiple categories.} 
%Anti-competition and Intellectual Property disputes accounted for the largest share of AI-related court opinions. 
%
While 39.77\% of relevant court opinions broadly referenced AI, others involved specific AI technologies, most notably Generative AI (12.87\%), predictive and decision-making systems (11.40\%), and virtual reality and vision technologies (8.77\%).
We also find the most common defendants were corporations (77.49\%), individuals (12.57\%), and government entities (7.02\%).
\item We found AI-related court opinions have more than doubled since 2023, primarily addressing disputes around AI through existing legal doctrines 
%(e.g., ranging from the Illinois Biometric Information Privacy Act (BIPA) to the Copyright Act of 1976)
. 
This reliance on existing laws has produced a form of ``piecemeal'' AI governance, where regulatory oversight emerges case by case rather than through a unified AI-specific framework, leading to some AI incidents remaining unresolved. 

\item We found disconnect between AI incidents and opinion outcomes, revealing gaps in coverage, definitions, and prevalence between documented and litigated harms. 

\item We make available an open-source dataset of the cases, metadata, as well as the qualitative codes and paper metadata at [Anonymized for review].
\end{enumerate}
%
%The cases litigated in federal courts are shaped more by which statutes are applicable, as opposed to where harms have occurred. 

%Overall, this work presents an in-depth mixed-methods investigation of AI-related federal court opinions in the US. We do not aim to be exhaustive across all possible jurisdictions, but rather aim to find an initial empirical foundation for a systematic analysis of legal arguments regarding AI incidents. By mapping these patterns, we show how litigation trends can inform better institutional design, more targeted regulation, and more realistic risk management. For policymakers, recurring disputes identify where new rules may be needed (or where enforcement bottlenecks make existing rights hard to vindicate). For AI developers, recurring allegations and successful defenses can guide governance priorities, such as data provenance, consent, and notice for sensitive identifiers, documentation of system capabilities and limitations, and careful management of AI-generated content used in legal and consumer-facing contexts.\\
%In summary, we contribute:

%\begin{enumerate}
 %   \item an analysis of 559 U.S. federal court opinions from 2006-2026 involving AI, resulting in a taxonomy of seven categories for legal contentions, six common AI systems litigated in court, and four common litigants as plaintiffs and defendants.
 %   \item an open-source dataset of the cases, metadata, as well as the qualitative codes and paper metadata, publicly available at [Anonymized for review].
%\end{enumerate}

\section{Related Work}
\paragraph{Surfacing AI Incidences}
Recent work across various research communities, such as AIES, STS, HCI have discussed and examined the harms of AI. For instance, \citet{weidinger2022taxonomy} developed a taxonomy of ethical and social risks associated with language models, including discrimination, information hazards, misinformation, malicious uses, HCI harms, and environmental harms. Similarly, \citet{bommasani2022opportunitiesrisksfoundationmodels} discussed the specific AI technologies and their associated societal impact such as inequity, misuse, economic and environmental impact.
Slattery et al. \cite{slattery2025airiskrepositorycomprehensive} tracked  AI incidences and offered a taxonomy.
Before generative AI, \citet{suresh2021framework} examined harms across the full machine learning life-cycle to anticipate, prevent and mitigate downstream unintended effects. %They analyzed different harms across multiple stages of the ML development pipeline, such as data collection, preparation, model development, evaluation, postprocessing, and deployment. 
Other work envisioned AI harms in different domains in detail, such as education~\cite{kasneci2023chatgpt}, environment~\cite{rillig2023risks}, healthcare~\cite{ali2023chatgpt}, marketing~\cite{kumar2024ethical}, finance~\cite{chen2024survey}, and medicine~\cite{gerke2020ethical, da2022legal}. 
Studies have also examined specific harms such as AI privacy violations \cite{10.1145/3613904.3642116, 10.1145/3768184.3768201}, discrimination in AI \cite{10.1613/jair.1.17806, 10.1145/3701613.3701615}, bias in medical applications of AI \cite{doi:10.1126/science.aax2342}, intellectual property (IP) disputes regarding generative AI \cite{ahuja2020artificial}, and potential human rights violation in AI systems~\cite{rodrigues2020legal}. 
Together, this literature offers broad coverage of potential harms grounded in technical analysis, normative argument, and lived experience.

Our work builds on this prior work but shifts the empirical lens from what harms are possible or anticipated to what harms are currently being  adjudicated. %Specifically, we analyze which AI-related harms are actually litigated and how they are framed, evaluated, and remedied in judicial reasoning. Court opinions not only reflect allegations of harm, but also the evidence, doctrines, and procedural constraints that shape which claims can be recognized by court. Our work connects the existing AI harm taxonomies to the issues that courts currently make visible.

\paragraph{Understanding Judicial Opinions}
% papers: 
Systematic coding of judicial opinions has been used to understand legal patterns with empirical methods. \citet{alexander2016misclassification} and \citet{barnett2018administrative} both reviewed large quantities of court opinions, resulting in empirical findings such as plaintiffs types in discrimination cases, or panel ideology's impacts on the win-rate of cases using Chevron Deference. More broadly, this approach builds on a long tradition of hand-coded judicial datasets, including the U.S. Supreme Court Database \cite{spaeth2014supreme}, which encodes issue areas, legal provisions, and case outcomes to facilitate large-scale quantitative analysis. Other work extends these methods, studying judicial behavior across all levels of the federal judiciary using structured empirical data \cite{epstein2013behavior}. 
Large-scale coding efforts have also been applied to track the judicial reception of specific legal texts, most notably the American Law Institute's Restatement project \cite{ali_restatements}. It represents one of the largest systematic efforts to collect and synthesize judicial decisions in American legal history, involving hundreds of judges, practitioners, and legal scholars over decades to manually review and classify case law across every major area of common law. %This effort has not incorporated AI-assisted methods, relying entirely on costly and time-intensive manual review.
Recent research has used computational methods such as machine learning and natural language processing to classify case outcomes, detect legal issues, or estimate ideological signals directly from text \cite{ash2017innovation, bommarito2018lexnlp}. We extend this work by using a mixed-methods approach to analyzing federal court opinions.%These approaches enable analysis at larger scale. 
%
%Together, this literature highlights both the value and the challenges of transforming unstructured judicial opinions into structured data for empirical legal analysis. 
%

%Our work takes a step toward addressing the resource-heavy challenges of extracting structured information from judicial opinions by exploring the use of LLMs. We follow a potential pathway to augment large-scale manual coding efforts like the ALI Restatement project, possibly reducing some of their cost and time burden.

\paragraph{AI Research and Policy Connections}
In AI governance scholarship, institutions determine how risks are defined, what accountability mechanisms exist, and which interventions are feasible~\cite{calo2017artificial, casper2025pitfalls}. For instance, \citet{calo2017artificial} mapped speculated AI capabilities to recurring policy problem areas (e.g., fairness/justice, safety, privacy/power, labor displacement) and to cross-cutting institutional questions about oversight capacity and accountability, such as the lack of governmental expertise required to properly regulate AI.
Recent work in HCI went further by empirically studying how AI governance is practiced and instrumented. For example, \citet{krafft2021action} studied how communities and local governments navigate automated decision-making and surveillance as public policy. \citet{Metcalf2021algorithm} examined how tools such as algorithmic impact assessments translate diffuse sociotechnical harms into actionable policies. \citet{kaushal2024automated} analyzed how regulatory regimes shape disclosure and accountability in platform contexts. In parallel, researchers have examined organizational governance programs~\cite{batool2025ai}, national initiatives such as Canada’s~\cite{attard2024governance}, and cross-sector ethics and accountability efforts~\cite{schiff2021ai}, amid rapidly evolving regulations including the EU AI Act.

However, much of this AI governance work treated law as a future policy target, rather than as an empirical site where AI harms are already being governed. \citet{atkinson2024legal} attempted to connect AI capabilities to \textit{anticipated} legal disputes and policy gaps. Instead, our work grounds the policy conversation in ``law-in-action'' by showing which harms become legally cognizable. %, who brings forth the harms, and how they do so. %This may highlight where existing legal institutions already function as a governance mechanism and where mismatches between harm landscapes and actionable claims suggest the need for new regulatory or organizational interventions.

\section{Methods}
To investigate AI-related court cases, we conducted a systematic analysis of AI-related federal court opinions in the United States. Opinions are written decisions issued by judges, providing the legal reasoning behind rulings. These documents contain factual background, procedural history, legal analysis, and references to precedent. %Reviewing the original court opinions allowed us to directly analyze the themes in each case, rather than reviewing scholarly articles, as is typical in conventional literature reviews.

\subsection{Data}
\label{sec:data}

We assembled a corpus of U.S. federal court opinions through GovInfo\footnote{GovInfo is the official website of the U.S. government that houses government information.\url{https://www.govinfo.gov/}, Alternatives to GovInfo, such as PACER, were either less accessible or less informative.}'s API, %{see PACER, https://pacer.uscourts.gov/; Federal Judicial Center, https://www.fjc.gov/.}
 retrieving all court opinions that included the terms \texttt{artificial intelligence} and \texttt{machine learning}, with a final query conducted on April 25, 2026. %Our initial query returned a large body of opinions spanning multiple courts, many of which included irrelevant opinions. For example, a case primarily about education discrimination (i.e., \textit{Lin v. University of Nebraska}) only mentioned the keyword once in passing: ``Lin took tests in [...] Artificial Intelligence Track'' at the university, suggesting that the court opinion was irrelevant for answering our research questions. 
To exclude irrelevant cases, we required the search keyword to appear in the first ten pages. %This filter ensured that we captured cases where AI plays a meaningful role in the dispute, while excluding opinions where AI is mentioned only incidentally (e.g., in appendices or exhibits).
While GovInfo began comprehensively storing federal court opinions starting 2004, we only found relevant cases starting in 2006. Our final search revealed 727 court opinions that included the term ``artificial intelligence'' in the first ten pages.

One author read each case to remove those where AI is not central to the dispute, meeting regularly with other authors to discuss any uncertainty.% by qualitatively annotating the initial sample data. 
%One author read each of the 727 cases to check whether the case's core contention is around AI, and removed cases where AI was ultimately unrelated. The authors met regularly to discuss any uncertain cases, and disagreements were resolved via consensus.
%
%For example, in \textit{Montgomery Jr., Adams-II v. Anderson et al}, one of the attorneys is mentioned to ``specialize in discrimination, employment matters, and AI''. However, the case mainly concerns employment-based racial discrimination against the plaintiff and was, thus, deemed irrelevant.
%
This removed 168 cases that were irrelevant to our inquiry, leaving 559 court opinions.

\subsection{Analysis}
To analyze the 559 court opinions, we iterative developed a codebook informed by our research questions. %around common areas of dispute, technologies involved, litigants, and arguments used by litigants. 

\paragraph{Dispute Topics} For the categories of disputes topics, we guided our initial exploration using Calo's framework of ``key challenges that AI pose for policymakers''~\cite{calo2017artificial}, which includes ``Justice and Equity'', ``Use of Force'', ``Safety and Certification'', ``Privacy and Power'', and ``Taxation and Displacement of Labor''. Calo's categories set a roadmap for potential areas for policy, however these do not necessarily reflect current litigation issues. For instance, ``Safety and Certification'' is a category that uniquely predicts the need for certification for AI, however we found no litigation for this category, and courts are not equipped to create general certifications for AI safety.

Thus, to further adapt Calo's work to our use-case, we conducted five iterations of independently applying and updating the codebook, using a randomly selected set of 10 cases from our sample in each iteration. In each iteration, the research team read the cases and applied the revised category to them. After each set, the research team came together to refine, merge, and replace existing codes, add new codes, and resolve disagreements through consensus. 
Doing so, we did not find a significant number of cases under ``taxation and displacement of labor'' which, as Calo articulated, concerns ``AI [displacing] jobs by mastering tasks currently performed by people.'' This highlights a potential gap between theoretical expectations that may be far in the future and what is observed empirically in court disputes today.
After finalizing the codebook, we used GPT-5 to annotate the remaining cases (see Appendix~\ref{appendix:catas}). To ensure reliability, we manually reviewed a random subset of LLM-coded cases of another 50 cases. For these cases, two annotators manually applied the existing code to the cases.%and check the inter-annotator agreement (IRR) using Cohen’s kappa on GPT-5's annotation. 
The resulting inter-annotator agreement on GPT-4's annotations is $\kappa_{\text{Case Topic}}=0.86$ indicating almost perfect agreement~\cite{mchugh2012interrater}. %This approach allowed us to scale the analysis with a reasonable accuracy and will also enable adding new cases in the future. 

\paragraph{Common AI Systems, Litigants, and Arguments}
We developed a codebook for AI Systems, Litigants, and Arguments, in line with our analysis for \emph{Dispute Topics}. We began with initial categories on our own, since litigation tends to center on relatively recurring AI system types (e.g., generative AI, virtual reality, recommender systems) and institutional party roles (e.g., individuals, corporations, governments). 
In the same qualitative coding iterations (10 cases per iteration) as \emph{Dispute Topics}, we additionally noted the systems and litigants in each case to iteratively refine or add system and litigant categories. 
We then prompted GPT-5 with the final codebook to annotate the rest of the cases in our sample (see Appendix~\ref{appendix:catas}). We similarly evaluated the LLM annotation with the random set of 50 cases, reaching a high IRR of $\kappa_{\text{Breakdown}}=0.94$.%, indicating almost perfect agreement~\cite{mchugh2012interrater}.

%On the other hand, argument patterns are the specific doctrines that parties use in pleadings. 
%Note that argument patterns are different from the dispute topics: 
Whereas dispute topics capture the broader themes of a dispute, argument patterns identify the specific legal doctrines, causes of action, and defenses that litigants invoke to translate those themes into specific legal claims. For example, a plaintiff may rely on the Telephone Consumer Protection Act (TCPA) to advance a privacy-related theory, despite TCPA being a consumer protection statute.
Thus, identifying argument patterns requires closer qualitative analysis than assigning dispute topics. %However, conducting such in-depth analysis for each court opinion is impractical at best. 
We used GPT-5 to exact argument patterns for both plaintiffs and defendants from each case using court opinions source text, synthesizing these patterns and identifying common argument patterns used by litigants.
As arguments are highly context-dependent and not always described exhaustively in judicial opinions, we did not aim to quantify argument patterns the same way as common dispute topics, common AI systems, and litigants. Rather, we used an exploratory coding approach to capture plaintiff- and defendant-side arguments as represented in the opinion.

\paragraph{Litigant AI Stakeholder Relations}
To identify litigant AI stakeholder relationships, we take inspiration from the EU AI Act \cite{smuha2025regulation}, mapping ``provider'' to AI owner, ``deployer'' to AI user, and ``affected persons'' to AI subject \cite{biden2023eo14110, 10.1007/s10462-023-10420-8, 10.1145/3593013.3594050}. We applied these codes to prompt GPT-5 to annotate each litigant per court opinion in our sample (see Appendix~\ref{appendix:relation}), and evaluated the LLM annotation on 70 randomly selected court opinion annotation, reaching a high IRR of $\kappa_{\text{Relation}}=0.84$. %, indicating almost perfect agreement~\cite{mchugh2012interrater}.

\paragraph{Identifying AI Incidents}
We drew incidents from the AI Incident Database (AIID) \cite{mcgregor2021preventing}, snapshotted on March 16, 2026. AIID incidents are classified using the MIT AI Risk Repository taxonomy \cite{slattery2025airiskrepositorycomprehensive}. A single coder mapped MIT taxonomy labels to our categories by comparing category descriptions.

\begin{table*}[t]
\small
\begin{tabular}{c>{\raggedright\arraybackslash}p{4.5cm}>{\raggedright\arraybackslash}p{11.5cm}}
\toprule
& \textbf{Label} & \textbf{Definition} \\
\midrule
\multirow{12}{*}{\rotatebox[origin=c]{90}{\litiganttext{\textbf{Litigants}}}} &
\multicolumn{2}{l}{\textit{Who litigates AI-related disputes? (as plaintiff/defendants)}} \\
\cmidrule(l){2-3}
& \litigant{Corporate entities} (40.35\%/77.49\%) & Companies, businesses who own or use AI systems.\\
& \litigant{Individuals} (43.86\%/12.57\%) & People affected or using AI systems, including consumers and class-action groups.\\
& \litigant{Government entities} (6.14\%/7.02\%) & Government bodies and related regulatory officials overseeing compliance to legislation.\\
& \litigant{Investors and shareholders} (6.73\%/0.58\%) & Individuals or groups that invest in AI companies.\\
& \litigant{Others} (2.92\%/2.34\%) & Litigants with fewer than 5 occurrences, including educational institutions, honors societies, non-profits, and religious organizations.\\
\midrule
\multirow{18}{*}{\rotatebox[origin=c]{90}{\roletext{\textbf{AI Technology}}}} &
\multicolumn{2}{l}{\textit{What AI technologies are involved in legal disputes?}} \\
\cmidrule(l){2-3}
& \role{General AI Technology} (39.77\%) & Systems that are described as containing AI, but which are not elaborated on further.\\
& \role{Generative AI} (12.87\%) & Systems that generate content, such as image or text generation models.\\
& \role{Predictive and decision-making systems} (11.40\%) & Systems that allocate resources, make decisions, or make predictions.\\
& \role{Virtual reality and vision technologies} (8.77\%) & Systems that analyze images, or create virtual or augmented realities.\\
& \role{Biometric identification technologies} (7.31\%) & Systems that verify identity or health using biological signals, such as facial recognition systems.\\
& \role{Recommendation and ranking algorithms} (6.73\%) & Systems that recommend, moderate, or rank content, typically featured in social media platforms.\\
& \role{Others} (13.16\%) & Technologies that have less than 5\% occurrence rate in our analysis, including general medical AI, automated calling systems, transcription AI, autonomous vehicle AI, and AI chatbots.\\
\midrule
\multirow{10}{*}{\rotatebox[origin=c]{90}{\adtext{\textbf{Case Topics}}}} &
\multicolumn{2}{l}{\textit{What are the common topics of legal disputes?}} \\
\cmidrule(l){2-3}
& \ad{Intellectual Property} (28.79\%) & On breaching intellectual property protections in the development or use of AI systems.\\
& \ad{AI in Legal Proceedings} (27.08\%) & On using AI as a tool in judicial process.\\
& \ad{Consumer Protection} (18.34\%) & On deceptive, misleading, or conducting unfair business practices involving AI.\\
& \ad{Data Protection} (9.17\%) & On violating privacy and data protections in the development or use of AI systems.\\
& \ad{Harmful Conduct} (8.10\%) & On harms from negligent or intentional conduct resulting in distress or identifiable harms.\\
& \ad{Justice and Equity} (4.69\%) & On creating, reinforcing, or remedying discrimination or systemic bias with AI.\\
& \ad{Anti-competition} (3.84\%) & On anti-competitive conduct or unfair competition involving AI systems or companies.\\
\bottomrule
\end{tabular}
\caption{Litigant, AI Technology, and Case Topic breakdown. Topic percentages computed out of 469 cases (relevant cases).}
\label{table}
\end{table*}
\section{Results}
\begin{figure}[H]
\includegraphics[width=1\columnwidth]{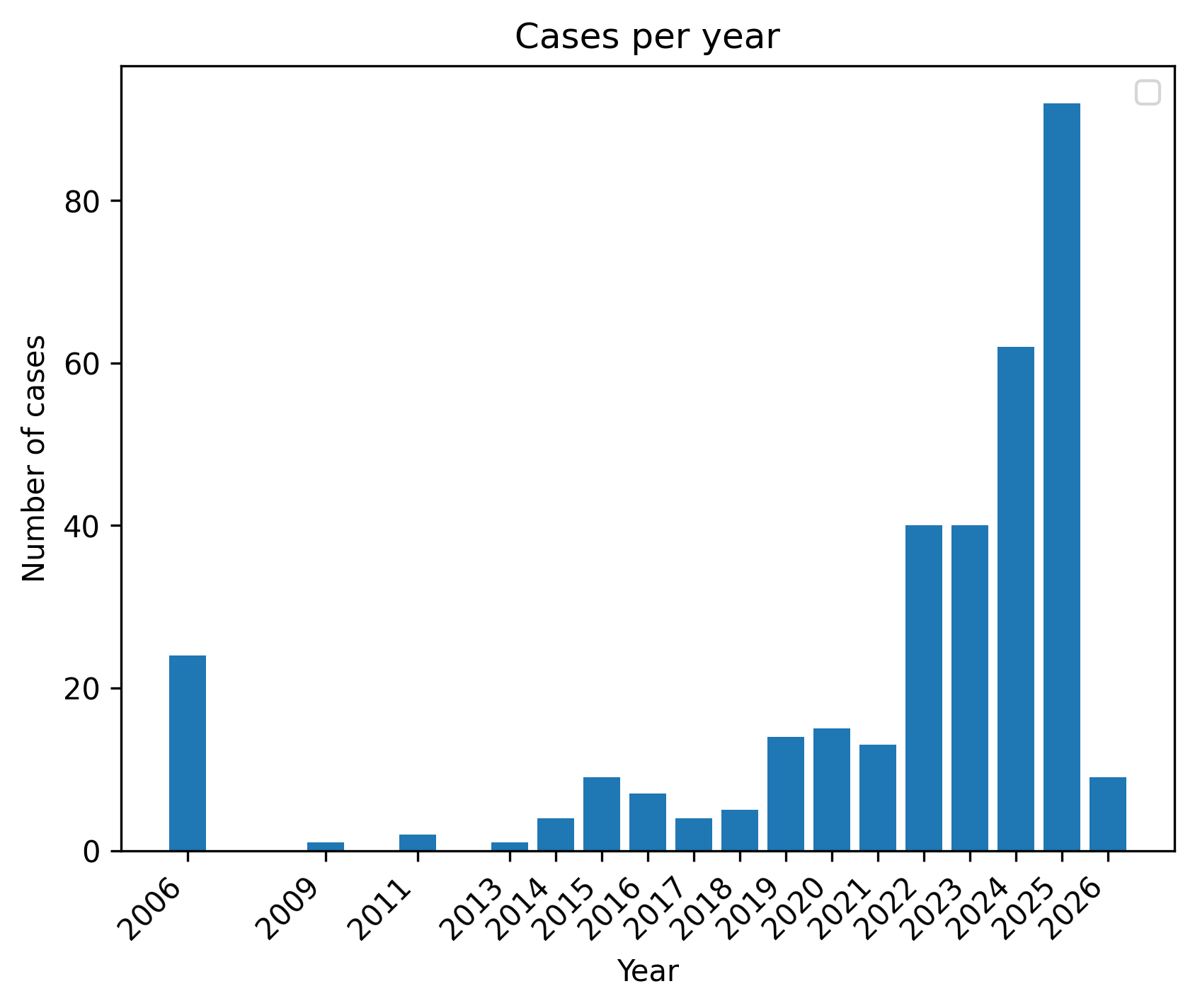}
\caption{Counts of AI-related court cases per year}
\label{appendix:raw-count}
\end{figure}

We first analyzed the AI court case distribution between 2006 (since GovInfo began recording them) and 2026. 
Figure 1~\ref{appendix:raw-count} shows that references
%References 
to AI were rare prior to the mid-2010s, followed by a substantial increase in the number of cases in more recent years, particularly after 2019.\footnote{In 2006, a cluster of cases against Quintus Corporation alleged fraud and misrepresentation related to ``artificial intelligence software,'' though the opinions did not elaborate on the role of AI.} Since 2023, the number of AI-related court cases more than doubled, showing increasing efforts to curtail competition and risks. A detailed breakdown of the specific technologies (see 
Appendix~\ref{appendix:figure1} and Table~\ref{table}) further shows a steady increase in litigation around generative AI. 

We identified four main litigants within the court opinions, six categories of AI technologies they litigate on, and seven common topics of their disputes (see Table~\ref{table}). %We found that cases most often occurred between corporations regarding intellectual property. 

\subsection{Who litigates AI-related disputes?}

AI-related litigation spans a range of actors including corporations, individuals, shareholders, and government entities. Our analysis focused on plaintiffs and defendants whose claims directly concern the development, deployment, or use of AI, excluding \ailegal cases \footnote{Litigant and tech counts exclude \ailegal as we investigate AI's wider impacts, not its literal legal application}.  

\textbf{\litigantlabel{Corporate entities}} are the most frequent actors both as plaintiffs and defendants ($N_\text{plaintiffs}=138$, $N_\text{defendants}=265$,). As plaintiffs, corporations most often bring claims related to \ip ($58.70\%$), reflecting efforts to protect AI-related data, patents, copyrights, and trademarks. 
In 41 cases, corporations alleged consumer protection claims through breach of business contract or unfair business practices involving AI-enabled products or services.
On the other hand, corporations are even more frequent defendants ($N=265$). Corporate defendants face claims across a wide range of legal areas, most commonly \consumerprotection ($N=118$) and \ip ($N=109$), but also \tort ($N=37$), \dataprotection ($N=45$), and \justice ($N=19$). This pattern reflects the central role of private firms in developing, deploying, and marketing AI systems, as well as the tendency for AI-related harms be attributed to corporate actors.
For example, in \case{Mullen Industries LLC v. Meta Platforms Inc.}, the plaintiff alleged that Meta infringed patents related to AI-enabled technologies, illustrating firms' reliance on existing IP law to contest ownership.
%
%The plaintiff alleged Meta's AR/VR headsets infringed upon their patents on AI-mediated control of virtual characters and augmented reality interfaces. Meta sought dismissal on the grounds that the plaintiff's claims relied on implausibly broad understandings of the patents, such as when the plaintiff assumed unspecified software was AI. Both plaintiff and defendant leveraged flexibility in interpretation of patent law to protect their respective technologies.

Under this broad category, it is worth noting the difference between asset-holding firms and platform-based firms. Asset-holding firms litigate primarily to protect or contest ownership interests, which appear most often as plaintiffs and rely heavily on IP law or contract claims. The AI systems in these cases are typically described in terms of patented methods, proprietary training data, or licensed technologies, and the alleged harms involve unauthorized copying, use, or disclosure. %Litigation in this category resembles earlier waves of technology disputes, with AI framed as an extension of existing commercial and informational assets rather than as a direct source of user-facing harm.
Platform-based firms (e.g. social media platforms, content-sharing services, and large-scale digital intermediaries) also fall under corporations, but appear in AI litigation under different conditions. We identified $27$ cases%\footnote{platforms are included in the $N_\text{defendants}$ count for corporations}  
in which platform companies were sued over the operation of AI-driven systems, particularly recommendation, ranking, and content moderation algorithms. These firms appear predominantly as defendants. Often these cases are ambiguous with the AI models, reflecting findings in Table~\ref{table}.

\textbf{\litigantlabel{Individuals}} typically lost ownership or control over AI systems and instead seek redress for downstream harms associated with AI deployment $(N_\text{plaintiffs} = 150, N_\text{defendants} = 43)$.
Their claims most often rely on \consumerprotection statutes ($45.33\%$), \dataprotection laws ($28.00\%$), and \tort claims ($22.67\%$). \ip claims ($27.33\%$) appear mainly when individuals allege misuse of creative works or personal data.

A common way for individuals to engage with AI-related lawsuits is forming a class-action group ($N=37$).\footnote{class-actions  are included in the $N_\text{plaintiffs}$ count for individuals} In U.S. civil litigation, a class action is a procedural mechanism that allows many individuals with similar claims to bring a lawsuit collectively rather than filing separate cases, not a separate legal entity.
%A class action is not a separate legal entity, rather it is multiple plaintiffs with common claims represented together in a single proceeding.
%
These cases most frequently rely on consumer protection statutes ($N=25$), with additional use of privacy ($N=16$) and tort claims ($N=9$). In \case{Taylor v. ConverseNow Technologies Inc.}, for example, a class of plaintiffs alleged that the defendant’s use of automated calling systems violated the Telephone Consumer Protection Act, allowing collective litigation to aggregate harms that would be difficult to pursue individually. These cases show how class actions surface AI-related harms that would otherwise fail on procedural or economic grounds.

\textbf{\litigantlabel{Government entities}} play several roles in AI litigation $(N_{\text{plaintiffs}} = 21, N_\text{defendants} = 24)$.
In \consumerprotection cases ($N=9$), the Securities and Exchange Commission (SEC) functions as a plaintiff when alleging securities fraud against companies whose main products involve AI. This features in cases such as \case{Securities and Exchange Commission v. Clayton et al.} and \case{Securities and Exchange Commission v. Tadrus et al.}, in which the SEC alleged misrepresentations regarding AI capabilities. %The government prosecuted criminal wire-fraud cases using AI in cases such as \case{USA v. Soto et al.} and \case{USA v. Sims et al.}. Furthermore, criminal cases such as \case{USA v. Thompson III} and \case{USA v. Linker} show the government prosecuting individuals who used generative AI to create child sexual abuse material. 
As defendants, government entities appear in eight \ip cases. For example, a series of cases brought by Stephen Thaler against the U.S. Patent and Trademark Office contested the agency’s refusal to recognize AI systems as inventors. Although limited in number, these disputes illustrate the courts' ability to review public-sector decision-making as new AI technologies impact present definitions. %Government actors also occasionally appear as plaintiffs in enforcement related to misleading AI claims, though such cases are less frequent in our sample. In other cases, such as \case{Centech Group Inc. v. USA} and \case{Association of American Universities et al. v. National Science Foundation et al.}, claims alleging unfair treatment regarding grants and contracts on AI may be brought against government entities.

\textbf{\litigantlabel{Investors and shareholders}} $(N_\text{plaintiffs} = 23, N_\text{defendants} = 2)$ 
concerns how AI-related misrepresentations affected investment decisions, with approximately $14$ cases alleging securities fraud or misleading capability claims. Such litigation extends AI accountability to financial markets by enforcing disclosure obligations and limiting exaggerated claims.
This is seen in \case{In Re Cerence Stockholder Derivative Litigation}, where shareholders alleged the defendant made misleading statements regarding their ``artificial-intelligence powered virtual assistants''.

\textit{\textbf{Others:}} The remaining categories are educational institutions (i.e., schools and universities), honors societies, non-profit organizations, and religions organizations, each of which have seven or fewer opinions.

\subsection{What AI technologies appear in legal disputes?}
\label{sec:ai-tech}

In our analysis, we found that legal contentions often occur around five main recurring AI technologies: generative models, biometric identification systems, predictive and decision-making systems, virtual reality and vision technologies and recommendation algorithms (see Table~\ref{table} 
%and frequency chart~\ref{appendix:figure1}
). However, we found that the majority of the cases, particularly those in \ip, \antitrust, and \consumerprotection, did not specify the exact AI technology used. We labeled these cases as the additional category: \textbf{\general} ($N=136$). 
For example, in \case{Pritchard v. Thompson et al.}, the discussion of patent infringement never continues beyond the vague description of the patented technology being ``AI-enabled dash cameras''. %In \case{WEX Inc v. HP In et al.}, the core rational for a trademark dispute is similarly discussed using broad language such as ``AI-driven data insights''.

\textbf{\genai} includes both generative language models and image generation models ($N=44$) \footnote{This does not include the \ailegal cases, which would predominantly belong to this category}. These cases frequently reference named text or image systems (e.g., ChatGPT, Claude, Llama, Stable Diffusion, Midjourney). The quantity of these opinions have increased, with the most opinions in 2025 ($N=25$).
For example, Claude is mentioned in an IP case regarding lyrics (\case{Concord Music Group Inc. et al v. Anthropic PBC}), ChatGPT is referenced in \case{New York Times v. OpenAI} (an IP case regarding written works), and \case{Andersen et al v. Stability AI Ltd.} discusses stable diffusion and the images created by artists.

\textbf{\decision} form the second most frequent category ($N=39$). These systems allocate risk, eligibility, or resources. Their influence on high-stakes outcomes invites legal challenges despite their opacity. An example is the MiDAS system in \case{Cahoo et al. v. SAS Analytics Inc. et al.}, an automated ``logic tree'' that  flags possibly fraudulent unemployment claims, which ultimately led to allegations of due process and civil-rights violations stemming from its use.% Similarly, in \case{The Estate of Gene B. Lokken, et al. v. UnitedHealth Group Inc.}, a class-action lawsuit was conducted due to the ``artificial intelligence models'' used by the defendant to determine whether medical insurance claims were necessary.

\textbf{\compvis} encompass image-analysis, virtual reality, and related imaging and vision technology ($N=30$). These systems are distinct from \bio as they do not involve the collection or storage of biometric identifiers, and are often litigated without statutes such as BIPA. Instead, when discussing \dataprotection cases, plaintiffs often frame their claims around privacy, surveillance, and data-use harms. For example, in \case{Ogletree v. Cleveland State University}, computer-vision techniques used to monitor a student’s testing environment prompted allegations of Fourth Amendment violations. %In related \ip cases, questions concerning the use of visual data to train AI models also arise, as in \case{UAB Planner5D v. Meta Platforms Inc. et al.}, where plaintiffs asserted copyright infringement tied to datasets used to develop scene-recognition systems.

\textbf{\bio}, particularly facial recognition systems appears less frequently but consistently across cases ($N=25$). This AI technology primarily involves privacy legal issues ($N=23$). These technologies are designed to establish or verify identity using biological signals, most commonly facial features. Unlike generative models, biometric systems do not produce expressive content but instead assign identity or similarity scores. 
Some examples include 
\case{Meehan v. VIPKid et al.}, where the facial geometry of teachers were taken without permission to train facial recognition systems, 
\case{Moomaw et al v. Geosnapshot Pty Ltd et al.}, where an online photo platform used facial recognition to identify individuals facial geometry and sell their data, 
and \case{Brightex Bio-Photonics LLC v. L' Oreal USA Inc.}, where a patent dispute occurred over facial recognition-based systems that use AI to identify skin blemishes.

\textbf{\rec} is typically referenced at the class-action level, with limited use of product-specific names, and are described using terms such as ``recommendation algorithm'' or ``algorithmic ranking system'' ($N=23$), providing scores or suggestions. We observed that these issues' bias (end results) are often under contention, but the exact AI models behind it are not often specified. For instance, Youtube's ``artificial intelligence and algorithms'' were the point of contention in \case{Divino Group LLC et al. v. Google LLC et al.}, where the algorithm allegedly discriminated against the plaintiff's content by censoring the plaintiffs due to their sexual orientation.

\textbf{Others:} The remaining categories are generic medical AI, automated calling systems, transcription AI, autonomous vehicle AI, and generic-purpose AI chatbots, each of which comprise less than $5\%$ of the opinions.

%%%%%%%%%%%%%%%%%%%%%%%%%%%%%%%%%%%%%%%%%%%%%%%%%%%%%%%%%%%%%%%%%%%%%%%%%%%%%%%%%%%%%%%%%%%%%%%%%%%%%%%%%%%%%%%%%%%%%%%%%%%%%%%%%%%%%%%%%%%%%%%%%%%%%%%%%%%%%%%%%%%%%%%%%%%%%%%%
\subsection{What are the common topics of legal disputes?}
Our analysis reveals that opinions regarding \ip and \consumerprotection are most common, with \ailegal showing AI as a litigation tool.

\textbf{\ailegal} ($38.82\%$, $N=217$): This label denotes cases where AI systems are merely used in the court processes, legal case management, or litigation tools. 
%The core contention is not about AI, but AI tools being used in the litigation process.
For example, in \case{Nonnie Berg v. United Airlines Inc}, the plaintiff ``improperly [used] artificial intelligence in generating the responses''  to the defendant's motion to strike the plaintiff's complaint. Most \ailegal cases follow this structure, with a party submitting AI-generated documents which receive judicial scrutiny. More recently, courts have addressed AI use in protective orders governing discovery, permitting litigants to use closed-source legal AI tools while explicitly forbidding the uploading of confidential information to open-source platforms that train on user data. \case{McCafferty v. University of Nebraska Medical Center et al} reflects this, where the court entered a stipulated protective order embodying exactly these limits.

\textbf{\ip} ($24.15\%$, $N=135$): This label denotes cases where the defendant allegedly violated copyrights, trademarks, trade secrets, or patents for AI technologies. These disputes include claims involving using copyrighted creative works for AI training or generation, infringing upon patented AI-enabled systems and methods, divulging trade secrets, violating trademarks on or with AI systems, and litigating concerns over the ownership and patentability of AI-generated outputs.

First, the \ip category includes a large body of cases about copyright infringement, particularly in disputes concerning training data and generative outputs. These cases typically involve allegations that defendants unlawfully used protected creative works (e.g., text, music, or visual art) to train generative AI systems. For example, in \case{Kadrey et al. v. Meta Platforms Inc. et al.}, the defendant allegedly illegally copied and used the plaintiff's copyright-protected written works to train the LLaMA model.
%
%Similar copyright disputes feature prominently in cases such as \case{Concord Music Group Inc. et al. v. Anthropic PBC}, \case{Andersen et al. v. Stability AI Ltd. et al.} and consolidated proceedings such as \case{In re Mosaic LLM Litigation}, \case{In re Google Generative AI Copyright Litigation}, and \case{In re OpenAI Inc. Copyright Infringement Litigation}.

In addition to copyright infringement claims, a small subset of cases address the copyrightability of AI-generated works. In cases such as \case{Stephen Thaler v. Shira Perlmutter et al.} and \case{Thaler v. Vidal}, the plaintiffs argued that AI-generated works should be eligible for copyright protection, notwithstanding the absence of direct human authorship.

Second, the \ip category also encompasses a large body of patent infringement cases that test the scope of patent protection for AI-enabled inventions.
For instance, in \case{VB Assets LLC v. Amazon.com Services LLC}, VB Assets, the patent holder, claimed that Amazon's Echo devices practiced the patented methods of its voice-based interaction, resulting in smart-speaker patent violation. 
%
%We found other patent infringement disputes concern patented methods, such as for classifying aerial imagery (\case{Aon Re Inc. v. Zesty.ai}), AI-assisted ear-worn devices (\case{Staton Techiya LLC v. Harman International Industries Incorporated et al.}), and voice-enabled systems for interacting with internet-based services (\case{Parus Holdings Inc. v. Apple Inc.}).

Third, the \ip category protects many trade secrets regarding AI, resulting in a large body of cases regarding the misuse or disclosure of trade secrets protected by non-compete agreements, non-disclosure agreements, and similar contracts. 
An example is \case{Nuance Communications Inc. v. Kovalenko}, where the defendant allegedly violated a non-compete agreement and illegally disclosed trade secrets regarding AI-powered radiology products. 
%
%Similarly, in \case{Admiin Inc. v. Kohan et al.}, the defendant allegedly used confidential information regarding AI marketplaces, harming the defendant's business.

A small body of cases investigate whether certain AI system implementations violated trademarks held by the plaintiffs. 
For instance, in \case{Overjet Inc. v. VideaHealth Inc}, both litigants offer dental AI products, however the defendant allegedly used confusingly similar branding to the plaintiff, violating the plaintiff's trademark.
%
%In \case{Groma LLC v. BuildRE LLC et al.}, the defendant allegedly used the plaintiff's trademarked branding for the defendant's AI products.

\textbf{\consumerprotection} ($15.39\%$, $N=86$): 
This label denotes cases where deceptive practices, unfair business practices, or misleading marketing of products occur with AI or regarding AI systems.
Unlike cases in categories such as \tort, \dataprotection, or \ip, \consumerprotection cases focus on market-facing harms, specifically deceptive, unfair, or misleading representations made to consumers, investors, or the public. Under this label, the claims are often about whether AI systems distort consumer choice, rather than whether the systems directly cause personal, privacy, or equity-based harms.
For example, in \case{Shift4 Payments LLC et al. v. JaredIsaacmanCourtCase.com et al.}, the defendants allegedly used AI-generated content to fabricate websites that violated the Anti-Cybersquatting Consumer Protection Act (ACPA), misleading users through false representations and associations. 
%
%Similarly, in \case{Federal Trade Commission (FTC) v. Automators LLC et al.}, the FTC alleged that the defendants misrepresented their ability to use AI and machine-learning tools to maximize customer revenues, constituting deceptive practices under consumer protection law.
%
Comparable allegations appear in \case{Federal Trade Commission v. Click Profit LLC et al.}, where the court found that the defendants made false or unsubstantiated claims regarding their purported use of AI to identify profitable products for online sales, in violation of the FTC Act. In these cases, AI was central not because of its technical operation, but because of how its capabilities are represented to consumers.
Consumer protection also captures securities fraud relating to AI companies in cases such as \case{D Agostino v. Innodata Inc.}, where the defendant allegedly issued false or misleading statements regarding the capability of their AI technology.

\textbf{\dataprotection} ($7.69\%$, $N=43$): This label denotes cases that frequently involve the collection, usage and storage of sensitive personal data without adequate notice or consent in AI systems. 
Many of these disputes arise under biometric privacy statutes, most notably Illinois's Biometric Information Privacy Act (BIPA), which allows for individuals to sue to protect their biometric data (e.g. facial geometry, fingerprints).
For instance, in \case{Dzananovic v. Bumble Inc.}, the plaintiff alleged the user authentication process tracks the user's facial geometry without users' explicit consent, in violation of BIPA. 
In \case{Lewis v. Maverick Transportation LLC et al.}, defendants allegedly obtained facial geometry from video recordings to use AI to monitor employee conduct. %Similarly, in \case{Tibbs et al. v. Arlo Technologies Inc.} facial geometry data was allegedly collected and stored without consent when the plaintiffs walked by the defendant's product (house-monitoring camera) during their delivery service work; and in both \case{Trio v. Turing Video Inc.} and \case{Naughton v. Amazon.com Inc.} facial geometries were allegedly stored after the plaintiff participated in COVID-19 screening software storing data obtained. All the BIPA cases involve litigation regarding the usage of facial geometry or private health data to train AI.

\textbf{\tort} ($6.80\%$, $N=38$): This label denotes cases where tort law, or similarly identifiable harms, provides the primary basis for liability in disputes involving AI systems. They center on allegations of legally cognizable harms, such as negligence, emotional or physical distress, or intentionally harmful conduct from using AI systems.
This includes both harms found in civil courts, as well as criminal charges related to harmful conduct arising from the use of AI systems.
Several cases illustrate tort as the central theory of liability. In \case{Edward Mansfield and Matthew Marchionnda, v. Norfolk Southern Railway Company}, plaintiffs brought forwards negligence claim from the defendant's use of AI systems to monitor rail-track conditions. According to the complaint, the defendant relied on automated assessments of rail-track health instead of physical safety measures, such as fencing, and this reliance contributed to foreseeable risks of harm. 
In \case{VoterLabs Inc. v. Ethos Group Consulting Services LLC}, VoterLabs Inc. alleged that the defendants used the plantiff's AI service without compensating them, framing the resulting financial loss as a tort-based negligence.

\textbf{\antitrust} ($3.22\%$, $N=18$): This category includes cases in which defendants are accused of having engaged in anti-competitive conduct related to AI technologies; that is, business practices that improperly restrict competition, exclude rivals, or maintain market power through means other than fair competition. Such conduct often involves claims of monopolization, attempted monopolization, or unlawful restraints of trade, particularly in the context of dominant technology platforms and artificial intelligence companies.
For example, in \case{Klein et al. v. Meta Platforms Inc.}, Meta allegedly attempted monopolization and restriction of trade by integrating AI models across dominant products in the social media and advertising sectors.
%
%Another example is \case{GovernmentGPT Incorporated et al v. Axon Enterprise Incorporated et al}, where GovernmentGPT ``is an early-stage startup that develops and markets artificial intelligence technologies, such as 360-degree body camera vests''. This case discusses ``a series of alleged illegal acquisitions and anti-competitive agreements occurring between [Axon, Microsoft and Safariland].''

\textbf{\justice} ($3.94\%$, $N=22$): This label denotes cases involving discrimination, bias, civil-rights or equal-protection violations, fairness or equity concerns, or systemic bias caused by or related to AI systems. These cases frame AI as a mechanism that produces, amplifies, or institutionalizes inequitable outcomes in employment, content moderation, and access to services.
For example, in \case{Brown v. Port Authority Transit Corporation et al.}, the plaintiff alleged that the defendants' use of AI-assisted communication tools in internal employment processes reflected racially discriminatory practices against Black employees (i.e., ``color and race are being used to systematically repress employment''). We found many discriminatory cases (e.g., on the basis of race, gender and age) in AI-assisted recruiting (see \case{Saas v. Major, Lindsey \& Africa LLC} and \case{EEOC v. iTutorGroup, Inc.}). 
Content moderation was found to also contain many cases regarding discrimination. In \case{Divino Group LLC et al. v. Google LLC et al.}, the plaintiffs (a group of LGBTQ{+} creators, ranging from psychologists to movie directors) alleged Google's AI-driven content moderation had systematically falsely flagged their content as obscene and sexually explicit, discriminating against LGTQ{+} creators.

%On the other hand, we found cases that advocate for the use of AI as a tool to promote justice and equity. In \case{In re Pinterest Derivative Litigation}, Pinterest shareholders claimed that the Pinterest leadership failed to address systematic workplace discrimination, resulting in a settlement that required Pinterest to adopt AI and data-driven tools to monitor and improve equity in hiring and promotion practices to mitigate discrimination. 

\subsubsection{Intersections across Case Categories}

Litigants often use multiple argument types to strengthen their case. %We thus found that cases often span multiple categories. 
We found that cases may be associated with multiple categories, showing co-occurrence frequency by examples.

First, we found that cases labeled under \ip may overlap with \antitrust. This overlap typically arises when defendants respond to IP claims by invoking competition-based arguments. 
For example, in \case{Thomson Reuters Enterprise Centre GmbH et al. v. ROSS Intelligence Inc.}, the plaintiffs alleged copyright infringement based on the defendants’ scraping of proprietary legal content to train an AI-based service. 
%
%In response, defendants justified their conduct by stating that their access to the plaintiff's data was necessary to promote competition, and therefore their use constituted fair use. This case, and similar cases, display how anti-competition arguments can be used to counter IP challenges in accessing AI systems or training data.
In response, defendants argued their access was necessary to promote competition and thus constituted fair use, illustrating how antitrust arguments can be leveraged to counter IP claims over AI training data.

A second common pattern of overlap arises between \dataprotection, \tort, and \consumerprotection. In these cases, the dispute centers on privacy, with claims brought under tort law, consumer protection law, and data protection statutes.
For example, in \case{Frasco v. Flo Health}, plaintiffs alleged their health data was sold without authorization. While the case implicates data protection statutes (i.e., California Confidentiality of Medical Information Act), the plaintiffs also asserted tort claims to address harms associated with loss of control over personal information. 

%Tort harms may be paired with consumer protection statues to better quantify harms experienced by the plaintiff. In \case{Black v. First Impression Interactive Inc}, the defendants used AI to generate the voice used in automated phone calls to contact the plaintiff. The plaintiff therefore claimed tort-based invasion-of-privacy harms and violations of the Telephone Consumer Protection Act (TCPA).
Tort harms may also be paired with consumer protection statutes, as in \case{Black v. First Impression Interactive Inc}, where the defendant used AI-generated voice in automated calls, prompting the plaintiff to claim both tort-based invasion of privacy and TCPA violations.
\case{Citizens Insurance Company of America v. Wynndalco Enterprises LLC et al}, labeled as \dataprotection, discussed tortious harms of invasion of privacy and privacy torts (with reference to alleged BIPA violations involving the defendant scraping and storing the of facial geometry of the plaintiffs).
While not as widespread as \tort in cases labeled by other categories, \consumerprotection policies can be seen in labels such as \dataprotection, such as when TCPA is used to litigate \case{Callier v. Freedom Forever Texas LLC} and \case{Perrong v. QuoteWizard.com LLC}, where consumer protection is used to protect privacy laws. However, the focus in TCPA cases can vary from privacy to consumer protection, as seen in \case{Mantha v. Quotewizard.com LLC}, where the focus is on the TCPA violation when being called by an automatic dialing system, as opposed to privacy harms. 
The intersection between \dataprotection and \consumerprotection is not limited to TCPA. In \case{In re TikTok, Inc., Consumer Privacy}, the defendant lost the case due to two consumer protection statutes: California Unfair Competition Law (UCL) which prohibits unfair, unlawful, or fraudulent business practices, and California False Advertising Law (FAL), which prohibits “unfair, deceptive, untrue, or misleading advertising”.

\subsection{Litigant Stakeholder Relation by Legal Category}

Per topics of legal disputes, we identified stakeholders types that litigate against each other
%, seen in Chart ~\ref{appendix:staketocat}
.

\paragraph{Intellectual Property ($n=135$)} AI owners were most often suing other AI owners ($n=57$), likely due to intellectual property infringement claims, where both owners have individual AI products, but one AI product infringes upon the other owner's intellectual property. An example is \case{WEX INC v. HP INC et al}, in which both parties own competing AI products. The plaintiff alleges that the defendant's AI product design infringed on the plaintiff's trademark.
For similar reasons, many AI owners may sue AI users who take the owner's intellectual property for their own use ($n=39$). This occurs in \case{Samsara Inc. v. Motive Technologies Inc}, where the plaintiff alleged that the defendant covertly accessed its platform through fake customer accounts to copy the plaintiff's patented AI dashcam technology.
AI subjects also sue AI owners when their intellectual property is used to train the AI owner's AI ($n=29$). \case{Concord Music Group Inc. et al v. Anthropic PBC} exemplifies this, where the plaintiff's musical lyrics were taken without their consent to train the defendant's large language model.
All other combinations of stakeholder relations had five or less court opinions.

\paragraph{Consumer Protection ($n=86$)} AI subjects were found to sue AI owners ($n=46$) or users ($n=48$) the most often in this category. AI subjects may sue AI owners for misrepresentation of product, such as in \case{City of Sunrise Firefighters' Pension Fund v. Oracle Corporation et al} where plaintiffs alleged the defendants used many misleading statements regarding the AI powered autonomous database when marketing their product. AI users may also misrepresent their application of AI, as seen in \case{Federal Trade Commission v. Empire Holdings Group, LLC}, where the defendant allegedly misrepresented that they used AI to maximize customer revenue.
AI subjects may also sue AI users when their use of AI violates consumer protection policies, such as in \case{League of Women Voters of New Hampshire et al v. Kramer et al}, where the defendant allegedly used AI to send out robocalls using Joe Biden's voice, violating the Telephone Consumer Protection Act.
%
%AI owners sue other AI owners as well ($n=6$), possibly due to defendant AI owner misrepresenting their technologies capabilities. This happens in \case{Highline Innovation Investment Partnership, LLC v. Biolert, LTD. et al}, where the plaintiff AI owner alleged that the defendant AI owner misrepresented the capabilities of their seizure detection technology, falsely claiming it used artificial intelligence and machine learning, to induce Highline to acquire Biolert's intellectual property.
All other combinations of stakeholder relations had at most six court opinions.

\paragraph{Data Protection ($n=43$)} AI subjects were found to primarily sue both AI owners ($n=26$) and AI users ($n=14$) for similar reasons. AI users were sued in \case{Ogletree v. Cleveland State University}, where the university's use of an AI proctoring system prompted a student's allegations of privacy invasion. In \textit{Gutierrez v. Wemagine.AI LLP}, the AI owner allegedly unconsensually stored the plaintiff's facial geometry after plaintiffs used the defendant's AI image altering website. Both incidents regard privacy violations, where either the AI user or AI owner stored data AI subject data without consent. All other combinations of stakeholder relations had at most one court opinion.

\paragraph{Harmful Conduct ($n=38$)} AI subjects were found to sue AI owners ($n=15$) and AI users ($n=12$). AI users were sued by AI subjects when they used AI to harm others, such as in \case{Phi Theta Kappa v. Honor Society}, where the Honor Society used AI to mislead the plaintiffs. Similarly, when the AI owner's technology creates or allows for harm, they are sued, such as in \case{In re Uber Technologies, Inc., Passenger Sexual Assault Litigation}, where Uber's machine learning model flagged a passenger's ride as high-risk for sexual assault but failed to act on that assessment by reassigning the driver or warning the passenger, resulting in the assault occurring. All other combinations of stakeholder relations had at most three court opinions.

\paragraph{Justice and Equity ($n=22$)} Primarily, AI subjects sue AI users ($n=14$) and AI owners ($n=7$). For example, in \case{Neuhtah Opiotennione v. Bozzuto Management Company}, the plaintiff sued the defendant for discriminating against older Facebook users by leveraging Facebook's algorithmic ad-targeting system to restrict housing advertisements to users aged 50 and under. AI subjects also sued AI owners ($n=7$), such as in \case{Mobley v. Workday Inc}, where the plaintiff alleged the defendant embedded an AI hiring system biased against candidates over 40 into multiple employers' pipelines. All other combinations of stakeholder relations had two or less court opinions.

\paragraph{Anti-competition ($n=18$)} AI owners sued AI users ($n=9$), typically in an attempt to argue that certain contracts regarding AI were unfairly rewarded to other parties, such as in \case{GovernmentGPT Incorporated et al v. Axon Enterprise Incorporated et al}, where the plaintiff alleged that the defendants violated federal antitrust law when one defendant (Safariland) sold a competing body camera company (VieVu) to the other defendant (Axon), thereby reducing competition in the market for police body cameras. %Similarly, in \case{EKAGRA PARTNERS, LLC et al. v. USA}, plaintiffs alleged the government's caps on a large business mentor's contributions to a joint venture proposal unlawfully restricted competition.
All other combinations of stakeholder relations had two or less court opinions.

\subsection{Incident and Court Opinion Disconnect}
Our findings also reveal gaps between academic AI risk taxonomies and litigated harms. To illustrate this gap, we draw from \citet{slattery2025airiskrepositorycomprehensive}'s taxonomy and \citet{mcgregor2021preventing}'s AI incident database to compare recorded AI incidents against court opinions in our findings. 
%
%We find the incident types and their frequencies in the repository do not align with current court opinion types. 

\textit{Categories for incidents are not necessarily reflected in court opinions.} For example, \ip (28.79\% of court opinions) has no direct parallel for incident categories, with \textit{Economic and cultural devaluation of human effort} (N$_{incident}$=2) being the closest parallel at only 0.20\% of the incidents.
Categories and subcategories such as \textit{AI system safety, failures, and limitations} (N$_{incident}$=261, 26.42\%) and \textit{Disinformation, surveillance, and influence at scale} \\(N$_{incident}$=84, 8.50\%) also have no clear counterpart, yet comprise a significant portion of AI incidents.

\textit{Court opinion and incident categories are sometimes definitionally misaligned.}
A single incident, such as ``creating humiliating or sexual imagery'' (as seen in \textit{Fraud, scams, and targeted manipulation} (N$_{incident}$=193, 19.53\%), may simultaneously draw on \tort, \consumerprotection, and \dataprotection to litigate. When comparing \citet{slattery2025airiskrepositorycomprehensive}'s \textit{Privacy \& Security} (N$_{incident}$=76, 7.69\%) category to \dataprotection, we notice definitions also vary greatly, with ``loss of confidential intellectual property'' bringing themes of \ip, and ``infer private information about individuals without their consent'' mentioning harms not often seen in \dataprotection .

Furthermore, \textit{category prevalence may be unequal between incidents and cases}. When proportionally compared, sub-categories \textit{Unfair discrimination and misrepresentation} (N$_{incident}$=115) and \textit{Unequal performance across groups} (n=32) (total N$_{incident}$=147, or 14.88\%) are more prevalent in AI incidents compared to court opinions about similar topics in \justice (4.69\%).

\section{Discussion}

%Between 2006-2026, litigation has been dominated by \ip and \consumerprotection disputes, largely involving corporate actors. \general is the most common technology, but cases recently involve more specific technologies (e.g., \genai) rather than the general reference to AI.
%
%Although public debate often spotlights harms such as bias and deepfakes under \justice, those risks appear only about 4.69\% of all AI-related opinions. %
Across the results, we see federal court outcomes being less about harms caused by AI, and more about whether a harm is cognizable under existing statues.
In this section, we discuss what this implies for AI governance.

\subsection{Courts are used for AI governance}
A central takeaway is that AI governance is operationalized through existing laws spanning multiple domains rather than bespoke AI statutes, historically applied to regulate emerging information technologies \cite{lyndon1995tort, ohlhausen2015competition}. Our categories correspond with doctrines drawn from these bodies of law, consistent with \citet{atkinson2024legal}, who identified similar doctrines as mechanisms for generative AI governance.

This breadth has two practical implications. First, it suggests the legal system addresses AI through existing tools rather than treating AI as a distinct legal object with unified rules, forming ``piecemeal'' governance \citet{tutt2017fda} warns against. The potential lack of legislation directly addressing harms caused by AI technology may lead to victims struggling to find legal recourse when those harms occur.
Second, it explains why litigation may be simultaneously salient and fragmented: plaintiffs must translate sociotechnical claims, such as training data provenance or discrimination in automated screening, into actionable legal theories, each with doctrinal thresholds, evidentiary burdens, and defenses.

When translating sociotechnical claims to legal theories, we see that AI-related claims often do not fit within just one legal category.
For example, antitrust theories may be raised to justify conduct challenged as copyright infringement, and privacy claims are justified using tort theories and consumer protection frameworks. 
This exemplifies how AI accountability is pursued: litigants often plead multiple theories to survive early motion practice, or to reach different remedies. This multi-theory approach is especially likely where harms are diffuse (privacy, discrimination) or technically complex, and where procedural barriers (standing, jurisdiction, arbitration) can prevent merits adjudication.

\subsection{Courts only capture part of the AI risk landscape}
Certain AI incident categories having no court opinion parallel mirrors the limitations we encountered when attempting to apply Calo's taxonomy to current court opinions. 

The misalignment reflects a deeper structural asymmetry: incident taxonomies are organized around harms experienced and the technology itself, whereas legal categories are organized around causes of action and doctrinal frameworks. 

The uneven distributions between incidents and cases likely reflects multiple compounding barriers. 
Constitutional equal protection doctrine and Title VII jurisprudence often require plaintiffs to demonstrate discriminatory intent rather than mere disparate impact (Washington v. Davis, 426 U.S. 229 (1976)), a burden that is exceptionally difficult to meet when the discriminatory mechanism is embedded in an opaque algorithmic system, as argued by \citet{10.1145/3617694.3623248}. 
Algorithmic discrimination harms are also often diffuse and individualized in ways that complicate class certification, limiting the procedural mechanisms available to aggregate small-scale injuries into viable litigation.

A potential reason for uneven prevalence and definitions is that certain incidents and harms may struggle to have a clear legal counterpart (e.g. \textit{Disinformation, surveillance, and influence at scale}), thus potentially leading to a lack of litigation regarding the harms, indicating a disconnect between the harms experienced by the public and the issues resolved in federal courts.
We also posit that certain AI technologies (e.g. \genai) have not been deployed widely or long enough to have many post-deployment incidents reach the federal court level.
Definitional disconnects, such as the one between \dataprotection incidents and \dataprotection court opinions, may signal the need for specific legislation that allows for harms to be more easily realized in the context of AI technology, as litigants currently may not have the legal tools necessary to realize harms in court. 

Furthermore, courts are structurally limited in their capacity to address AI risks proactively. As \citet{b0922b0b36004ef89f832abb847fac2e} note, tort liability for negligence typically requires that actual harm be realized. In the eyes of the law, risk alone is insufficient to establish a legal claim. The prevailing requirement of concrete injury means that AI incidents concerning risks or near-misses that have not yet created downstream harm to identifiable plaintiffs are unlikely to be litigated. \emph{This renders courts largely reactive as a governance mechanism, as they are poorly positioned to intervene preemptively in the face of emerging AI risks.}

\subsection{Challenges with procedural or doctrinal thresholds}
Many AI-related cases may be decided on whether the plaintiffs cleared procedural or doctrinal thresholds, without reaching the alleged impact caused by the AI-system.
These barriers operate at two levels. First, threshold barriers may prevent courts from reaching the merits entirely. Defendants frequently raise personal jurisdiction challenges, arguing insufficient forum contacts, or challenge standing by contending that plaintiffs failed to allege a concrete and particularized injury. Contractual defenses especially mandatory arbitration provisions serve as a recurrent mechanism to redirect disputes away from judicial adjudication altogether. These defenses can be outcome-determinative, particularly for cross-border or distributed AI services where the locus of conduct and harm is contested.

Second, substantive doctrinal barriers can defeat meritorious claims. Defendants invoke statutory immunities such as CDA § 230 to argue that claims are barred regardless of the AI system's technical operation. In discrimination cases, doctrinal thresholds requiring proof of discriminatory intent rather than demonstrating disparate impact can block claims even where algorithmic bias is well-documented, leaving some harms unlitigated \cite{10.1145/3715275.3732213}.

Platform liability cases illustrate this dynamic. In cases involving AI-driven recommendation and content moderation algorithms, defendants regularly invoke CDA § 230 to argue they are immune from liability as intermediaries of third-party content. When courts grant these motions, the question of whether the algorithm actually caused cognizable harm, for example, by systematically amplifying or suppressing content in discriminatory ways, is never reached. In \case{Force v. Facebook, Inc.}, the Second Circuit held that § 230 barred claims arising from Facebook's algorithmic recommendation of content linked to terrorist attacks, even though the plaintiffs' theory of harm centered on the algorithm's active role in content amplification rather than passive hosting. The ruling did not evaluate whether the algorithm caused harm; it held that the statutory framework foreclosed the inquiry. This pattern in which doctrinal barriers prevent merits adjudication of AI-related harms recurs across our dataset, particularly in cases involving recommendation systems.

For AI governance, this shows that rights on paper may not translate into rights in practice if plaintiffs cannot establish jurisdiction, standing, or avoid arbitration. For system builders, it also matters as many of these issues are shaped by business and product choices.%: forum selection clauses, arbitration terms, representations in product documentation, and the architecture of cross-border service provision.
To circumvent this, plaintiffs in
\ip and \dataprotection disputes often surface questions about upstream AI development and deployment pipelines, including the use of training data, ownership of AI-generated outputs, and biometric or sensitive data processing. The \ip label explicitly covers training data disputes for AI and ownership of AI-generated content. \dataprotection includes unauthorized data collection and biometric privacy claims, with BIPA appearing as a recurring vehicle for challenges involving facial geometry. These patterns suggest a key reality of AI accountability today: \emph{even when harms are experienced downstream, plaintiffs often pursue upstream leverage points} (data collection, disclosures, marketing representations) as existing law supplies clearer doctrinal hooks or statutory damages for those cases. 

This dynamic has significant governance implications. \emph{The legal system is incentivizing transparency and disclosure obligations over outcome-based accountability for AI-driven harms.} AI system opacity compounds this problem: plaintiffs must both clear procedural and doctrinal thresholds, while also rendering technically complex systems legible to judges and juries who may lack relevant expertise \cite{10.1145/3531146.3533084}, further reinforcing the tendency to pursue claims where the legal theory does not require explaining the AI system's internal operation.

\subsection{AI is reshaping the legal process itself}
The existence of an \ailegal category (27.08\%, N=217) indicates a second-order shift: AI is becoming part of the litigation workflow, not merely the object of litigation.
These cases often involve scrutiny of AI-generated filings or litigation conduct. However, there are also cases where AI is a proposed tool for litigation itself such as \case{Jonathan Bertuccelli and Studio 3, Inc. v. Universal City Studios LLC, et al.} where automated facial analysis was used to quantify perceived similarity in an IP dispute, and \case{Abdul Nevarez v. Forty Niners Football Company} where AI was used for discovery. Prior work also highlights the proposed use of AI in courts to increase accessibility to justice, or to provide users with legal advice \cite{avery2023chatgpt, chien2024generative, 10.1145/3630106.3659048}. 
These findings indicate that \emph{courts are developing norms around AI tool use as they enter briefs, pleadings, and discovery practices.}
%Judicial responses shape acceptable legal practice. 
%On the other hand, legal processes have become a site of AI literacy production. 
%Judges, clerks, and litigants are collectively deciding on how to evaluate AI-generated material, including what kinds of errors are tolerable and what kinds signal sanctionable conduct. 
%\citet{grossman2023gptjudge} notes this, stating courts must actively adapt to detect and regulate AI use in courts. 

\section{Limitations and Future Work}
Our findings should be interpreted in light of methodological constraints. The corpus is derived from GovInfo opinions retrieved via two search terms; this choice may have excluded cases where AI is salient but described using different terminology (e.g., ``algorithm'' or ``automated decision system'') or discussed later in an opinion. %The removal of 168 cases during the robustness check further illustrates that keyword-based approaches can still retain irrelevant opinions even after filtering.
In addition, our computational pipeline (i.e., the LLM-generated summaries and categorization) supports scale and consistency but introduces its own validity risks. These methods are best understood as exploratory and supportive, rather than as establishing ``natural'' categories of AI litigation.

More broadly, court opinions are informative but they only reflect what reaches adjudication and what results in written decisions. Nevertheless, opinions provide structured descriptions of asserted facts, legal theories, and judicial reasoning, making them valuable for understanding how AI-related conflicts are formalized and contested through law. Future work may consider analyses beyond court opinions from paid legal (but prohibitively expensive) repositories, such as WestLaw and LexisNexis, even at the state level.

\section{Conclusion}
Courts are increasingly asked to adjudicate disputes over how AI systems are built, trained, deployed, marketed, and used
%in consequential settings
. In this paper, we examined AI-related U.S. federal court opinions to provide an empirical snapshot of (1) what types of AI disputes appear in opinions, (2) which AI systems are the subject, (3) who brought AI-related lawsuits to court.
Our results showed that (1) litigation typically occurred between corporate parties regarding \ip and \consumerprotection; (2) AI-harms are litigated using pre-existing statutes 
%(e.g. BIPA, TCPA, Copyright Act) 
rather than AI-specific law, (3) procedural thresholds such as standing and arbitration often disposed of claims before harms could be litigated, and (4) 
several harm categories present in AI incident taxonomies are rarely litigated. %Discrimination appears at less than a third of its reported rate in \citet{slattery2025airiskrepositorycomprehensive}, and AI safety failures and disinformation at scale are largely absent from our corpus. Conversely, the largest litigation category found (\ip) has no direct parallel in the incident taxonomy.  
These findings suggest that federal court outcomes are shaped less by where AI causes harm, and more by which harms are cognizable under existing statutes.

\bibliography{aaai2026.bib}

\clearpage

\appendix
\section{Prompts}

\subsection{Prompt to Summarize Court Opinions}
\label{appendix:summary}
\rule{\columnwidth}{0.4pt}\\
\textbf{Summarizing Agent: Prompt to Summarize Opinions}\\
\texttt{
You are a legal expert. Please summarize the case: [Case Title].\\
Here is the context of the case: [Court opinion text].\\
Rule 1: Summarize the case in a few sentences.\\
Rule 2: If the case is related to artificial intelligence (AI) or machine learning (ML) or automated systems, please focus on how the AI is involved in the summary.\\
Rule 3: Respond with JSON: {"summary": "summary\_text"}\\}
\rule{\columnwidth}{0.4pt}

\subsection{Prompt to Extract Common AI Systems, Litigants, and Arguments}
\label{appendix:catas}
\rule{\columnwidth}{0.4pt}
\texttt{
You are extracting structured information from a legal case.\\
Read the ENTIRE case and return ONLY valid JSON that follows the schema exactly.\\
Do NOT include explanations, markdown, or any fields not listed in the schema.\\
If information is missing, use an empty string "" or empty array [] — do NOT guess.\\
The schema is as follows:}
\begin{verbatim}
{
    "case_id": {"type": "string"},
    "core_ai_system": {"type": "string"},
    "plaintiff_labels": {
        "type": "array", 
        "items": {
            "type": "string"
        }
    }, 
    "plaintiff": {
        "type": "object",
        "properties": {
            "name": {"type": "string"},
            "entity_type": 
                {"type": "string"}
        },
        "required": 
            ["name", "entity_type"]
    },
    "defendant_labels": {
        "type": "array", 
        "items": {
            "type": "string"
        }
    }, 
    "defendant": {
        "type": "object",
        "properties": {
            "name": {"type": "string"},
            "entity_type": 
                {"type": "string"}
        },
        "required": 
            ["name", "entity_type"]
    }
}
\end{verbatim}
\texttt{\\
CASE NAME: [Case Title]\\
SUMMARY: [Case Summary]
}
\begin{verbatim}
LABEL DESCRIPTIONS: {
1. Antitrust: market competition, 
monopolization, market power,
anti-competitive practices, market 
dominance, price-fixing, exclusive 
dealing, or restraint of trade involving
ANY tech companies, or anti-competitive
practices by major platforms or AI 
companies.
2. IP Law: patents, copyrights, 
trademarks for AI models or tech, or 
training data disputes, AI-generated
content ownership.
3. Privacy and Data Protection: data 
breaches, unauthorized data collection 
by automated systems, or privacy 
violations involving algorithms or data
processing.
4. Tort: physical harm, emotional 
distress, negligence, defamation, or 
personal injury involving ANY automated 
systems, tech systems, major tech 
corporations using AI, or algorithms.
5. Justice and Equity: discrimination, 
bias, civil rights violations, equal 
protection issues, equity issues, 
fairness concerns, or systemic bias, 
alleged or substantiated discrimination 
or bias caused by AI, automated systems, 
or algorithms, or related to AI, 
automated systems, or algorithms. (e.g., 
hiring, lending, search).
6. Consumer Protection: deceptive 
practices, unfair business practices with 
tech/automated systems, or misleading
marketing of tech products or AI 
capabilities.
7. AI in Legal Proceedings: AI systems 
are merely used in the court processes, 
legal case management, or litigation 
tools. The core contention is not about 
AI, but AI tools have been used in the
litigation process.
8. Unrelated: cases that have no 
meaningful connection to AI, ML, or 
automated systems. If the case is 
frivolous, or involves discrimination,
privacy, or other issues *without 
automation/AI/algorithmic involvement*, 
classify as Unrelated.
}

Follow instructions:
- Fill in the JSON fields only
- For case_id, use the case name/filename
- For core_ai_system: Read through the 
entire case text, extract the text 
describing where AI is mentioned, and the 
AI technology used.
- Given the label descriptions, write the 
label into the JSON if the litigant used 
an argument thatfits the description of 
the label.
- Be specific about outcomes
- For plaintiff / defendant: Describe the 
party, not just the literal name.
\end{verbatim}
\rule{\columnwidth}{0.4pt}

\subsection{Prompt to Identify Stakeholder Relations}
\label{appendix:relation}
\begin{verbatim}
AI_ROLE_SCHEMA = {
    "name": "ai_role_extraction",
    "schema": {
        "type": "object",
        "properties": {
            "plaintiff_ai_role": {
                "type": "string",
                "enum": ["AI user"
                    , "AI subject"
                    , "AI owner"],
            },
            "defendant_ai_role": {
                "type": "string",
                "enum": ["AI user"
                    , "AI subject"
                    , "AI owner"],
            },
        },
        "required": ["plaintiff_ai_role"
            , "defendant_ai_role"],
        "additionalProperties": False,
    },
    "strict": True,
}

SYSTEM_PROMPT = """You are a legal 
expert analyzing AI-related court 
cases. Return ONLY valid JSON matching
the schema. No explanations, markdown, 
or extra text."""

USER_PROMPT_TEMPLATE = """Classify the 
AI role(s) of the plaintiff and 
defendant in this court case.

Definitions:
- AI user   : the party uses AI as a 
tool in their business, products, or 
operations
- AI subject: the party is directly 
affected by or subjected to an AI 
system (e.g., their data was used for
training, they were profiled or
surveilled by AI, they suffered harm 
from an AI output)
- AI owner  : the party owns, develops, 
licenses, or holds intellectual 
property for an AI system or AI-based
technology

Choose only one of the 3 definitions.

CASE TITLE: {title}

PLAINTIFF: {plaintiff_raw}
DEFENDANT: {defendant_raw}

TECHNOLOGY AT ISSUE: {tech_raw}

CASE SUMMARY:
{summary}

Return JSON with keys 
"plaintiff_ai_role" and 
"defendant_ai_role", each a string 
with one value chosen from: "AI user", 
"AI subject", "AI owner"."""
\end{verbatim}

\section{Figures}
\begin{figure}[H]
\includegraphics[width=1\columnwidth]{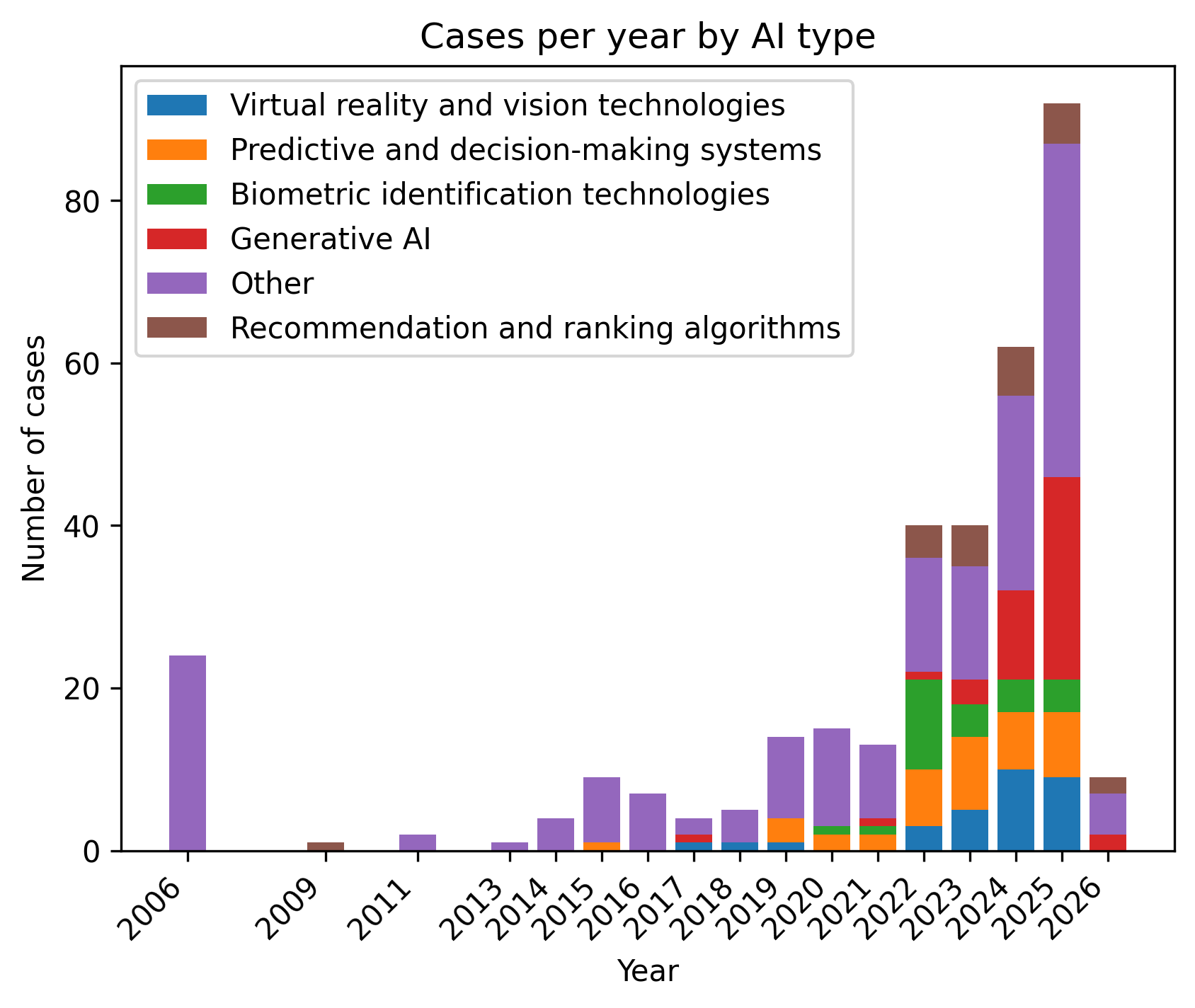}
\caption{Counts of AI-related court cases per year by AI technology category}
\label{appendix:figure1}
\end{figure}

\begin{figure}[H]
\includegraphics[width=1\columnwidth]{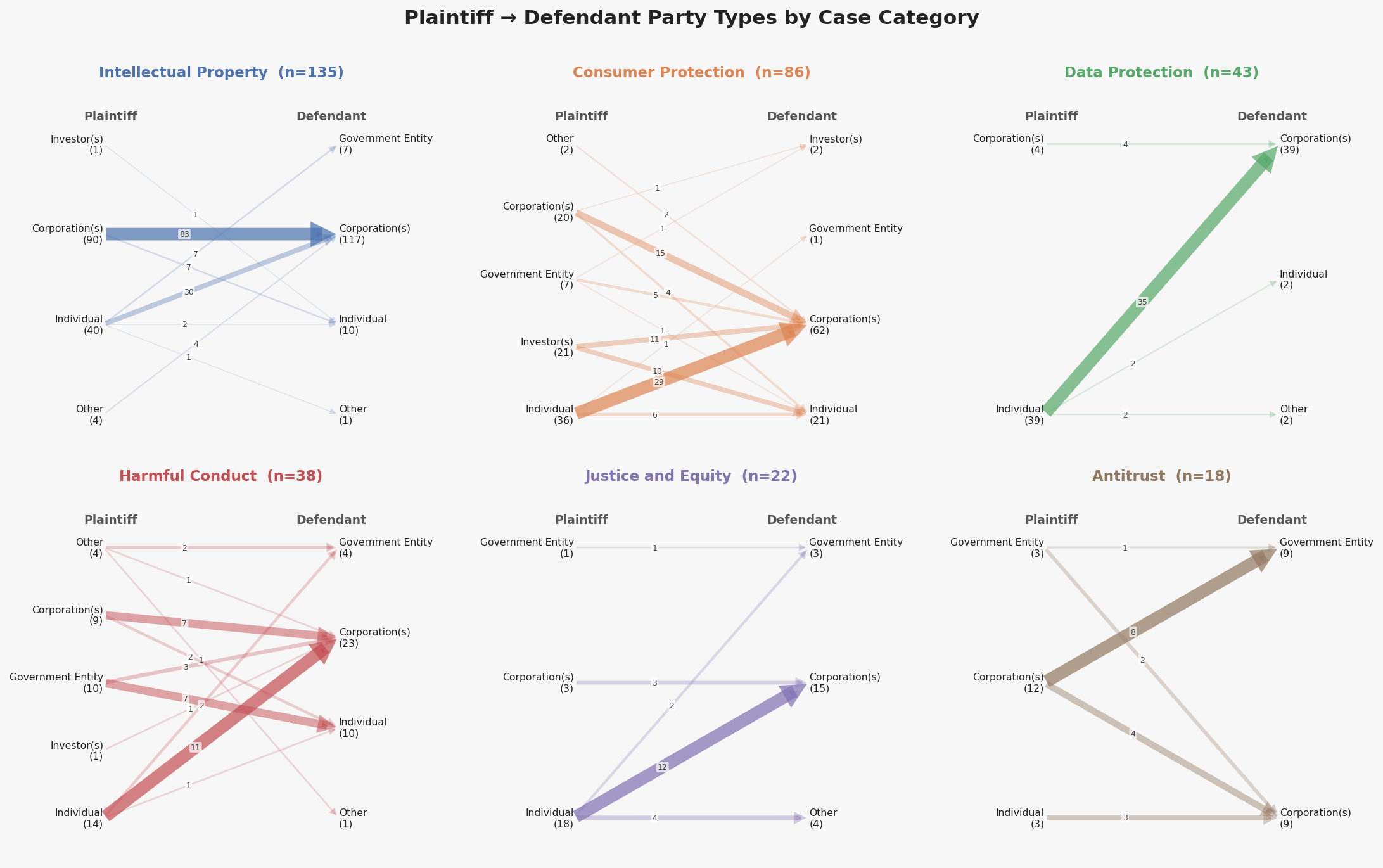}
\caption{Litigant conflict counts per opinion category}
\label{appendix:staketocat}
\end{figure}

\begin{figure}[H]
\centering
\includegraphics[width=1\columnwidth]{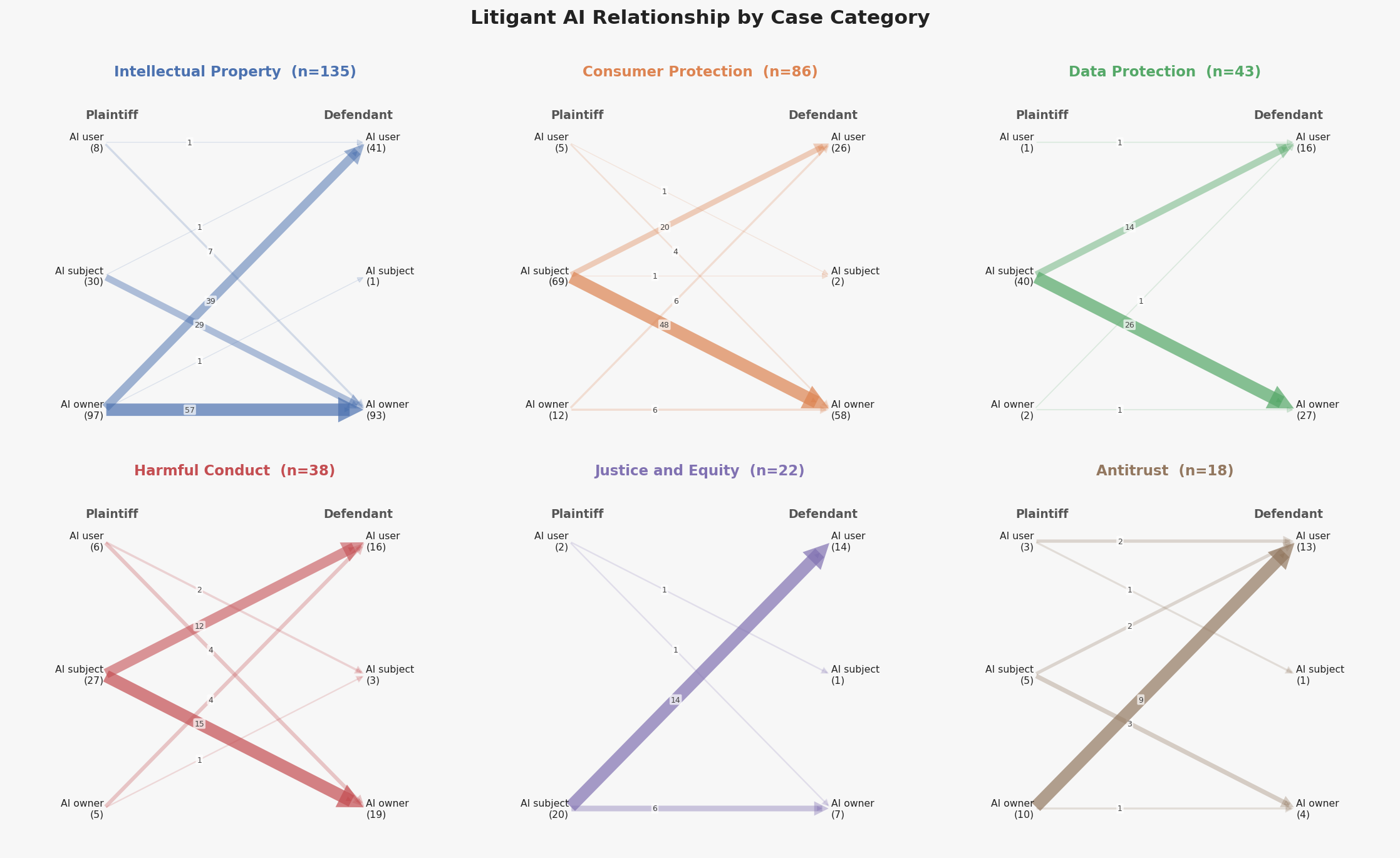}
\caption{Litigant conflict counts per litigant relation}
\label{appendix:relationtocat}
\end{figure}
\begin{figure}[H]
\centering
\includegraphics[width=1\columnwidth]{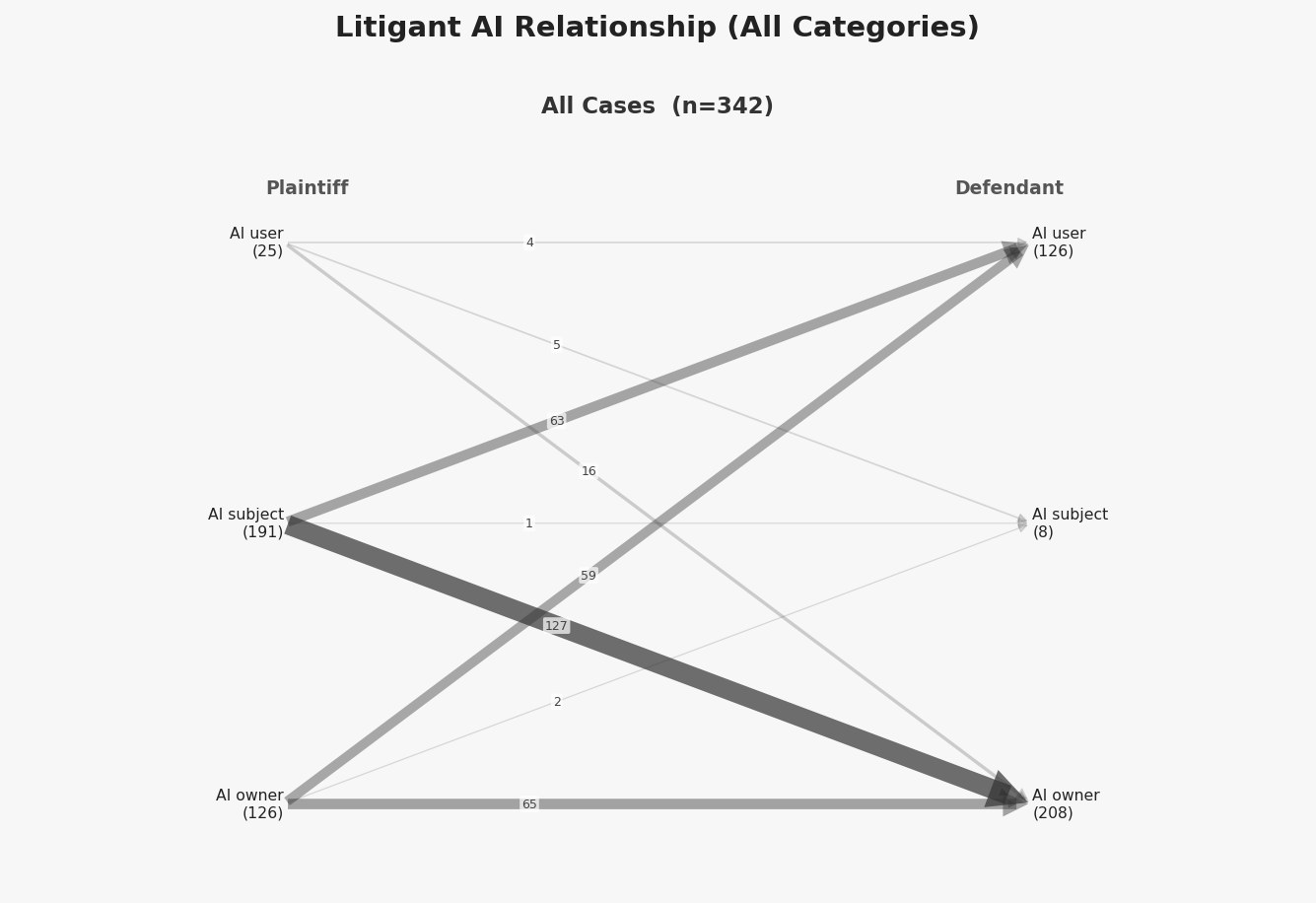}
\caption{Cumulative litigant conflict counts per litigant relation}
\label{appendix:system}
\end{figure}
\end{document}